\documentclass[aps,showpacs,prb,twocolumn,amsmath,amssymb]{revtex4}

\usepackage{amsmath}
\usepackage{amssymb}
\usepackage{graphicx}
\usepackage{dcolumn}
\usepackage{bm}

\begin{document}

\bibliographystyle{apsrev}

\newcommand{\boldtau}{{\bm \tau}}
\newcommand{\boldphi}{{\bm \phi}}
\newcommand{\hvarphi}{\hat{\varphi}}

\newcommand{\UR}{URu$_2$Si$_2$}
\newcommand{\CC}{CeCoIn$_5$}
\newcommand{\UP}{UPd$_2$Al$_3$}

\newcommand{\la}{\langle}
\newcommand{\ra}{\rangle}
\newcommand{\da}{\dagger}

\newcommand{\lc}{\lowercase}
\newcommand{\hg}{\hat{g}_0}

\newcommand{\no}{\nonumber}
\newcommand{\be}{\begin{equation}} 
\newcommand{\ee}{\end{equation}}
\newcommand{\bea}{\begin{eqnarray}} 
\newcommand{\eea}{\end{eqnarray}}

\newcommand{\vQ}{{\bf Q}}
\newcommand{\vk}{{\bf k}}
\newcommand{\vq}{{\bf q}}
\newcommand{\vM}{{\bf M}}
\newcommand{\vm}{{\bf m}}
\newcommand{\vH}{{\bf H}}
\newcommand{\vh}{{\bf h}}
\newcommand{\Q}{($0,0,\frac{2\pi}{c}$)}
\newcommand{\De}{$\Delta$({\bf k})}
\newcommand{\DeQ}{$\Delta$({\bf k}+{\bf Q})}
\newcommand{\Detr}{$\Delta({\bf k}+{\bf Q}) = - \Delta({\bf k})$}
\newcommand{\tth}{\tilde{T}_h}

\title{Signatures of hidden order symmetry in torque oscillations, elastic constant anomalies and field induced moments in URu$_2$Si$_2$}
\author{Peter Thalmeier$^1$ and Tetsuya Takimoto$^2$}
\affiliation{$^1$Max Planck Institute for Chemical Physics of Solids,
01187 Dresden, Germany\\
$^2$Asia Pacific Center for Theoretical Physics, Pohang University of Science
and Technology,\\
Pohang, Gyeongbuk 790-84, Korea}
\date{\today}
\begin{abstract}
We discuss the  conclusions on the symmetry of hidden order (HO) in URu$_2$Si$_2$ that may be drawn from recent torque experiments in rotating magnetic field by Okazaki et al. \cite{Okazaki11}.
They are very sensitive to changes in the magnetic susceptibility induced by HO.  We show that the observed twofold angular torque oscillations give evidence  that hidden order has degenerate E- type (yz,zx) symmetry where both components are realised. The oscillations have the wrong characteristics or are absent for the 1D nontrivial representations like quadrupolar $B_1 (x^2-y^2)$ and  $B_2 (xy)$ type HO or hexadecapolar $A_2(xy(x^2-y^2))$ type HO. Therefore they may be excluded as candidates for hidden order. We also predict the field-angular variation of possible field-induced Bragg peaks based on underlying E-type order parameter and discuss the expected elastic constant anomalies.
\end{abstract}


\maketitle


\section{Introduction}
\label{sect:introduction}

The moderate heavy fermion compound \UR~ serves as a paradigm of  electronic hidden order (HO) and field induced quantum phase transitions.
Although investigated by many techniques since considerable time its microscopic nature and in particular its symmetry remains elusive.
Progress in understanding was greatly advanced through pressure experiments which demonstrated phase separation of hidden order (HO)  phase and parasitic heterogeneous antiferromagnetism (AF) with tiny moments (see Ref.~\onlinecite{Amitsuka07} and references cited therein). At low pressure the bulk of high quality samples exhibits only HO with ordering temperature T$_h$=17 K  which coexists with the superconducting phase appearing below $T_c$ =1.3 K. The HO transition at $T_h$ is associated with a large entropy release that can be accounted for by a dramatic increase in the gap of incommensurate itinerant spin excitations \cite{Wiebe07}. In addition to ARPES \cite{Santander09} and STM \cite{Schmidt10} and transport \cite {Behnia04,Behnia05,Kasahara07} experiments  this shows that a microscopic model for the hidden order has to take into account the partly itinerant character of 5f electrons which has been discussed extensively in Ref.~\onlinecite{Oppeneer10}. The microscopic models considered sofar are too numerous to be discussed here (see references cited in Ref.~\onlinecite{Okazaki11}). The purpose of this work is a more modest one: To reconsider the question of proper symmetry class of the HO parameter from analyzing new oscillatory torque measurements \cite{Okazaki11} and previous elastic constant measurements \cite{Kuwahara97}. We will propose that the former lead to a unique answer for HO symmetry in \UR.\\

For the analysis of symmetry classes it is convenient to start from localised 5f electron states which can carry multipolar moments up to rank five in 5f angular momentum operators. This approach allows a systematic and physically appealing classification of  HO symmetry.
Each of those with rank $\geq 2$ may be candidate for hidden order because unlike the dipolar (rank 1) moments they are hidden in zero field neutron diffraction experiments without magnetic field. The full classification is given in Table~\ref{table:table1}.
Indeed the first serious discussion of HO symmetry in \UR~\cite{Santini94} used such localised picture. The tetragonal symmetry group $D_{4h}$ of \UR~ has three nontrivial 1D representations (A$_2$, B$_1$ and B$_2$) and one doubly degenerate E representation. Any possible multipolar HO has to belong to one of them denoted by $\Gamma$. Since pressure dependence shows that HO is in competition with AF order at the commensurate $\vQ = (0,0,1)$ wave vector (in bct structure this is equivalent to $\vQ =(1,0,0)$) one may expect the same (antiferro) ordering wave vector for HO. This is supported by the observation that the resonance excitation found at $\vQ$ by inelastic neutron scattering (INS) turns into the elastic AF peak above the critical pressure of HO \cite{Villaume08} and by the similarity of Fermi surfaces in the AF and HO phases \cite{Hassinger10}.
Furthermore the absence of any macroscopic lattice distortion \cite{Jeffries10} below T$_h$ is evidence against the ferro HO case with $\vQ = 0$. Nevertheless we will also include ferro-type HO in our analysis.
 Most frequently quadrupolar HO type (O$_\Gamma$) has been invoked. It was proposed to be of B$_1$(O$_{x^2-y^2}$) type  \cite{Ohkawa99} or  B$_2$(O$_{xy}$) type \cite{Santini94,Ohkawa99,Harima10}. In addition higher order multipole moments like octupoles of  $B_2$(T$_{xyz}$) type \cite{Santini94,Kiss05} and hexadecapolar order A$_2$(H$_{xy(x^2-y^2)}$) \cite{Santini94,Haule10} have been suggested and discussed. For a general review of multipolar type order in crystalline electric field (CEF) split f-electron systems see Refs.~\onlinecite{Santini09,Kuramoto09}.\\

 The most powerful means of identifying the HO consists in inducing dipolar (magnetic) order at the same wave vector by application of an external field. The induced magnetic order  can then be seen in neutron diffraction. The directions of induced moment as function of field orientation reflect the symmetry of the underlying hidden order which can then be identified. For example by this method the HO in some rare earth skutterudite compounds has been determined \cite{Kohgi03,Kuramoto09}. The HO should also lead to subtle changes in the uniform dipolar (magnetic) and quadrupolar susceptibilities. The latter can be probed with high sensitivity through the elastic constant anomalies below the HO transition. The magnetic susceptibility tensor in the HO phase has recently been investigated with the equally sensitive torque method \cite{Okazaki11} below T$_h$ which we will discuss first.
Above the HO transition the susceptibility tensor should have fourfold symmetry in the tetragonal ab plane corresponding to fourfould oscillations in the torque. Below T$_h$ this may no longer be true and twofold torque oscillations may appear depending on the symmetry of the HO. Their observation therefore allows to draw conclusions on the latter.
The torque exerted on the sample of volume V by an external field is given by $\boldtau=\mu_0V\vM\times\vH$ with $\vM=\tensor{\chi}\vH$  where $\tensor{\chi}$ is the susceptibility tensor.  For field rotation in the ab-plane with $ \vH =  H(\cos\varphi,\sin\varphi)$ the non-zero component
$\tau_z(\varphi)\equiv \tau(\varphi)$ is given by
\bea
\tau(\varphi)=\mu_0VH^2\Bigl[\frac{1}{2}(\chi_{xx}-\chi_{yy})\sin 2\varphi-\chi_{xy}\cos 2\varphi\Bigr]
\label{eq:torque1}
\eea
where $\varphi$ is counted from the a axis and $\chi_{xy}=\chi_{yx}$ is assumed here. To analyse the torque oscillations it is therefore necessary to compute the susceptibility tensor above and below the HO transition.\\ 

In Sec.~\ref{sect:multipoles} we first give a symmetry classification of possible multipole order parameters in \UR~ within a local f-f coupling scheme,  in analogy to Ref.~\onlinecite{Takimoto08}. In Sec.~\ref{sect:Ltorque} a Landau theory for selected even rank multipoles is used to calculate the torque oscillations. In Sec.~\ref{sect:dipole} we consider the possibility of field induced staggered moments in the HO phase and in Sec.~\ref{sect:elastic}  we discuss the anomalies of some symmetry elastic constants at the HO transition.
Finally Sec.~\ref{sect:conclusion} gives the conclusion.


\section{Symmetry classification of multipolar order parameters for \lc{f}-electron systems}
\label{sect:multipoles}
In $f$-electron systems, due to strong spin-orbit coupling, 
$f$-electron states are described well by the lower $j$=5/2 multiplet 
where $j$ is the total angular momentum. 
The multiplet consists of three Kramers doublets. 
Then, due to six states, 
irreducible tensors for representing multipoles are available up to rank 5. 
With  increasing rank {\it k} they are denoted as
dipole (J, rank 1), quadrupole (O, rank 2 ), octupole (T, rank 3) hexadecapole (H, rank 4), 
and dotriacontapole (D, rank 5) moments. The latter will not be included in the discussion and the tables.

The irreducible tensor has 2$k$+1 components $J^{(k)}_{q}\; (|q|\leq k)$, 
which satisfy the following relations
\begin{eqnarray}
  &&[J^{z}, J^{(k)}_{q}]=qJ^{(k)}_{q}\\
  &&[J^{\pm}, J^{(k)}_{q}]=\sqrt{(k\mp q)(k\pm q+1)}J^{(k)}_{q\pm 1}
\end{eqnarray}
where $J^{\pm}$=$J^{x}\pm{\rm i}J^{y}$ are raising and lowering operators, 
$J^{\alpha}$ is $\alpha$-component of total angular momentum 
operator. 
When these multipole moments are in a CEF produced by cubic or tetragonal surroundings, 
the corresponding tensors have to be classified according to 
irreducible representations of the point group \cite{Shiina97} which is $D_{4h}$ here.
It is convenient to express the multipole operators that act on CEF states by single particle operators adapted 
to the $D_{4h}$ symmetry. We start from the tetragonal CEF states of U in f-f coupling scheme given by
\begin{eqnarray}
  &&|\Gamma_7^{(1)},\pm\rangle=|\gamma=1,\pm \rangle=
   a|\pm\frac{5}{2}\rangle
   -\sqrt{1-a^2}|\mp\frac{3}{2}\rangle \no\\
  &&|\Gamma_7^{( 2)},\pm\rangle=|\gamma=2,\pm \rangle=
   \sqrt{1-a^2}|\pm\frac{5}{2}\rangle
   +a|\mp\frac{3}{2}\rangle \no\\
  &&|\Gamma_6,\pm\rangle=|\gamma=3,\pm\rangle=|\pm\frac{1}{2}\rangle 
\end{eqnarray}
where $a$ is a real number determined by the parameter set 
of tetragonal CEF. 
Furthermore,  $\sigma$ denotes pseudo-spin states of $f$-electrons 
in the Kramers doublets which are connected by time reversal and $\gamma$ 
the various doublets. Two Kramers doublets of $\gamma$=1 and $\gamma$=2 
correspond to $\Gamma_{7}$ irreducible representation in the tetragonal 
crystal structure 
while the remaining one with $\gamma$=3 belongs to 
$\Gamma_{6}$ irreducible representation. 

By using these CEF bases for $f$-electron states, 
we rewrite operators of irreducible tensors $J^{(k)}_{q}$. 
When $q$ is an even (odd) number, 
the matrix element 
$\langle\gamma\sigma|J^{(k)}_{q}|\gamma'\sigma'\rangle$ 
can only be non-zero for $\sigma$=$\sigma'$ ($\sigma$=$-\sigma'$)
Using this property 
irreducible tensors with $q$=$2p$ are given by 
\begin{eqnarray}
  J^{(k)}_{2p}=\sum_{\gamma,\gamma'}\sum_{\sigma}
              \langle\gamma\sigma|J^{(k)}_{2p}|\gamma'\sigma\rangle
              f_{\gamma\sigma}^{\dagger}f_{\gamma'\sigma}
\end{eqnarray}
where the matrix element 
$\langle\gamma\sigma|J^{(k)}_{2p}|\gamma'\sigma\rangle$ 
satisfies  
%
\begin{eqnarray}
  &&\langle\gamma\sigma|J^{(k)}_{2p}|\gamma'\sigma\rangle
  =(-)^{k}\langle\gamma'\bar{\sigma}|J^{(k)}_{2p}|\gamma\bar{\sigma}\rangle
  \nonumber\\
  &&\hspace{21mm}=\langle\gamma'\sigma|J^{(k)}_{-2p}|\gamma\sigma\rangle
\end{eqnarray}
Similarly, 
irreducible tensors with $q$=$2p+1$ are given by 
\begin{eqnarray}
  J^{(k)}_{2p+1}=\sum_{\gamma,\gamma'}\sum_{\sigma}
              \langle\gamma\sigma|J^{(k)}_{2p+1}|\gamma'\bar{\sigma}\rangle
              f_{\gamma\sigma}^{\dagger}f_{\gamma'\bar{\sigma}}
\end{eqnarray}
with  
%
\begin{eqnarray}
  &&\langle\gamma\sigma|J^{(k)}_{2p+1}|\gamma'\bar{\sigma}\rangle
  =(-)^{k+1}\langle\gamma'\sigma|J^{(k)}_{2p+1}|\gamma\bar{\sigma}\rangle
  \nonumber\\
  &&\hspace{26mm}
  =-\langle\gamma'\bar{\sigma}|J^{(k)}_{-(2p+1)}|\gamma\sigma\rangle
\end{eqnarray}
Using these expressions, 
all $J^{(k)}_{q}$ finally are given by simple linear combinations of 
charge operator $\rho_{\gamma\gamma'}$ 
and pseudo-spin operators 
$S^{\alpha}_{\gamma\gamma'}$ ($\alpha$=x, y, and z) defined by
\begin{eqnarray}
  \rho_{\gamma\gamma'}=\frac{1}{2}\sum_{\sigma}
      f_{\gamma\sigma}^{\dagger}f_{\gamma'\sigma},\hspace{5mm}
  S^{\alpha}_{\gamma\gamma'}=\frac{1}{2}\sum_{\sigma,\sigma'}
      f_{\gamma\sigma}^{\dagger}\tau^{\alpha}_{\sigma\sigma'}
      f_{\gamma'\sigma'}
      \label{eq:oneparticle}
\end{eqnarray} 
%
%
\begin{table}[t]
\begin{center}
\begin{tabular}{ccc}
\hline
$\Gamma$(D$_{\rm 4h}$) & $J^{(k)}_{q}$ & $\phi^{\Gamma}_{n}$\\
\hline
${\rm A}^{-}_{1}$ 
 & $D_4$
  & $\phi^{A^{-}_{1}}_{1}$=
    $\frac{\rm i}{\sqrt{2}}(\rho_{12}-\rho_{21})$\\
\hline
${\rm A}^{-}_{2}$ 
 & $J^{z}$ & $\phi^{A^{-}_{2}}_{1}$=$S^{z}_{11}$\\
 & $T^{\alpha}_z$ & $\phi^{A^{-}_{2}}_{2}$=$S^{z}_{22}$\\
 \hline
${\rm B}^{-}_{1}$ 
 & $T_{xyz}$
  & $\phi^{B^{-}_{1}}_{1}$=
    $\frac{\rm i}{\sqrt{2}}(\rho_{23}-\rho_{32})$\\
\hline
${\rm B}^{-}_{2}$ 
 & $T^{\beta}_z$
  & $\phi^{B^{-}_{2}}_{1}$=
    $\frac{1}{\sqrt{2}}(S^{z}_{23}+S^{z}_{32})$\\
\hline
${\rm E}^{-}$ 
 & $J^{x}$
  & $\phi^{E^{-}}_{1x}$=$S^{x}_{11}$\\
 & $T^{\alpha}_x$
  & $\phi^{E^{-}}_{2x}$=$S^{x}_{22}$\\
 & $T^{\beta}_x$
  & $\phi^{E^{-}}_{3x}$=$S^{x}_{33}$\\
 \cline{2-3}
 & $J^{y}$
  & $\phi^{E^{-}}_{1y}$=$S^{y}_{11}$\\
 & $T^{\alpha}_y$
  & $\phi^{E^{-}}_{2y}$=$S^{y}_{22}$\\
 & $T^{\beta}_y$
  & $\phi^{E^{-}}_{3y}$=$S^{y}_{33}$\\
\hline\hline
${\rm A}^{+}_{1}$ 
 & $N_{f}$ & $\phi^{A^{+}_{1}}_{1}$=$\rho_{11}$\\
 & $O_2^0$ & $\phi^{A^{+}_{1}}_{2}$=$\rho_{22}$\\
 & $H_0$ & $\phi^{A^{+}_{1}}_{3}$=$\rho_{33}$\\
 & $H_4$
  & $\phi^{A^{+}_{1}}_{4}$=
    $\frac{1}{\sqrt{2}}(\rho_{12}+\rho_{21})$\\
\hline
${\rm A}^{+}_{2}$ 
 & $H^{\alpha}_z$
  & $\phi^{A^{+}_{2}}_{1}$=
    $\frac{\rm i}{\sqrt{2}}(S^{z}_{12}-S^{z}_{21})$\\
\hline
${\rm B}^{+}_{1}$ 
 & $O_2^2$
  & $\phi^{B^{+}_{1}}_{1}$=
    $\frac{1}{\sqrt{2}}(\rho_{23}+\rho_{32})$\\
 & $H_2$
  & $\phi^{B^{+}_{1}}_{2}$=
    $\frac{1}{\sqrt{2}}(\rho_{31}+\rho_{13})$\\
\hline
${\rm B}^{+}_{2}$ 
 & $O_{xy}$
  & $\phi^{B^{+}_{2}}_{1}$=
    $\frac{\rm i}{\sqrt{2}}(S^{z}_{23}-S^{z}_{32})$\\
 & $H^{\beta}_z$
  & $\phi^{B^{+}_{2}}_{2}$=
    $\frac{\rm i}{\sqrt{2}}(S^{z}_{31}-S^{z}_{13})$\\
\hline
${\rm E}^{+}$ 
 & $O_{yz}$
  & $\phi^{E^{+}}_{1x}$=
    $\frac{\rm i}{\sqrt{2}}(S^{x}_{12}-S^{x}_{21})$\\
 & $H^{\alpha}_x$
  & $\phi^{E^{+}}_{2x}$=
    $\frac{\rm i}{\sqrt{2}}(S^{x}_{23}-S^{x}_{32})$\\
 & $H^{\beta}_x$
  & $\phi^{E^{+}}_{3x}$=
    $\frac{\rm i}{\sqrt{2}}(S^{x}_{31}-S^{x}_{13})$\\
 \cline{2-3}
 & $O_{zx}$
  & $\phi^{E^{+}}_{1y}$=
    $\frac{\rm i}{\sqrt{2}}(S^{y}_{12}-S^{y}_{21})$\\
 & $H^{\alpha}_y$
  & $\phi^{E^{+}}_{2y}$=
    $\frac{\rm i}{\sqrt{2}}(S^{y}_{23}-S^{y}_{32})$\\
 & $H^{\beta}_y$
  & $\phi^{E^{+}}_{3y}$=
    $\frac{\rm i}{\sqrt{2}}(S^{y}_{31}-S^{y}_{13})$\\
\hline
\end{tabular}
\end{center}
\caption{The first column shows the
irreducible representation of D$_{\rm 4h}$. 
The second and third ones describe corresponding bases 
consisting of multipole moments and their
one-particle operators (Eq.~(\ref{eq:oneparticle})), 
respectively. 
Here, in the second column of A$^{+}_{1}$ irreducible representation, 
$N_{f}$ is the number operator of $f$-electron. 
The superscript $\pm$ of irreducible representation 
expresses the parity with respect to time reversal.}
\label{table:table1}
\end{table}
%
where $\hat{\tau}^{\alpha}$ are the Pauli matrices and $f_{\gamma\sigma}^\dagger$ creates 
f electrons in CEF Kramers doublet state $|\gamma,\sigma\rangle$. 
These charge and pseudo-spin operators which are either symmetric 
or antisymmetric in the orbital space are convenient
as new bases of multipole moments. 
Thus, we introduce a new Hermitean basis $\{\phi^{\Gamma}_{n}\}$ 
which is shown in the third column of Table \ref{table:table1}. The hidden order parameter
must belong to one of these representations. In particular they  may be used to construct microscopic models for HO. Although this is not the topic of the present investigation we briefly describe the principal type of possible microscopic models. 

In primarily localised models the multipoles of CEF split \cite{Santini94,Kiss05} 5f states given in Table \ref{table:table1} are coupled by effective intersite interactions obtained from eliminating the conduction electrons, see e.g. Ref.~\onlinecite{Shiba99}. Depending on the leading interaction channel and appropriate symmetry of lowest CEF states a specific multipole order belonging to one of the representations in Table \ref{table:table1} may be selected as the ground state. Although this is a local picture it should be noted that staggered order of multipoles may have strong feedback on the itinerant states by gapping large parts of the Fermi surface. This mechanism has been invoked in Pr- skutterudites and has been proposed as similar in
nature to the observed carrier depletion in the HO phase of \UR~\cite{Kuramoto09}.

In the itinerant models the hidden order parameter results from a pairing mechanism of electrons and holes leading to an unconventional density wave or 'nematic' state. In such models the gap function (either in spin singlet or triplet channel) breaks the crystal symmetry and has a nontrivial (e.g. quadrupolar) type form factor or {\bf k}-dependence. These models have been frequently proposed as candidates for hidden order in itinerant systems, e.g. in Refs.~\onlinecite{Nersesyan91,Ikeda98}. They naturally lead to a partial gapping of the Fermi surface in the HO state. Such exotic pairing states can only be stabilized in extended Hubbard type models  for the itinerant 5f bands.

In this work we do not proceed to the construction of a microscopic model for HO but rather try to draw conclusions from a symmetry analysis of predicted torque oscillations based on Landau theory of HO.

\section{Landau theory of torque oscillations in the HO phase}
\label{sect:Ltorque}

As candidates for HO we consider only the even representations of Table \ref{table:table1} because rank 1 dipoles as primary order parameter are obviously excluded and rank 3 octupoles may induce dipoles as secondary order even at zero field \cite{Kiss05}. In particular we discuss rank 2 nondegenerate quadrupole order parameters $\phi =\phi_1^{B_1^+}$,  $\phi= \phi_1^{B_2^+}$ or twofold degenerate $(\phi_a,\phi_b)=(\phi_{1x}^{E^+},\phi_{1y}^{E^+})$. In addition we consider the rank 4 nondegenerate hexadecapolar order $\phi=\phi_1^{A_2^+}$.
The Landau free energy functional F contains the uniform magnetisation {\bf M} and the 
HO parameter. For the 1D representations it can be constructed from invariants of the D$_{4h}$ group using only terms up to fourth order in the HO and uniform magnetization:
\bea
\label{eq:fho1}
F^0_{HO}\{\phi\}&=&a\phi^2+\frac{1}{2}b\phi^4\no\\
B_1: F_{HO}\{\phi\}&=&F^0_{HO}\{\phi\}+\no\\
&&g\phi(M_x^2-M_y^2) +\tilde{g}\phi^2(M_x^2+M_y^2) \no\\ 
B_2: F_{HO}\{\phi\}&=&F^0_{HO}\{\phi\}+\no\\
&&g\phi M_xM_y +\tilde{g}\phi^2(M_x^2+M_y^2)\\
A_2: F_{HO}\{\phi\}&=&F^0_{HO}\{\phi\}+\no\\
&&g\phi M_xM_y(M_x^2-M_y^2)+\tilde{g}\phi^2(M_x^2+M_y^2) \no
\eea
Note that the terms $\sim$ g which are linear in the order parameter $\phi$ appear only for ferro- type HO and are strictly forbidden ($g_2\equiv 0$) for antiferro-type HO, e.g. $\vQ = (0,0,1)$ due to the requirement of translational invariance. In this case only the terms $\sim \phi^2$ contribute. Because for  all 1D HO the square of $\phi$ is an invariant it can only couple to the modulus of the magnetisation 
 which does not break the in-plane symmetry. The argument would also apply for even terms (the only ones allowed for $\vQ\neq 0$) of higher than fourth order.  This fact has important consequences for the possible torque oscillations in the 1D HO phases.\\
 
For the twofold degenerate E representation the free energy is 
\bea
\label{eq:fho2}
F^0_{HO}\{\phi_a,\phi_b\}&=&a(\phi_a^2+\phi_b^2)+\frac{1}{2}b_1(\phi_a^4+\phi_b^4) +b_2\phi_a^2\phi_b^2 \no\\
F_{HO}\{\phi_a,\phi_b\}&=&F^0_{HO}\{\phi_a,\phi_b\}+g_0(\phi_a^2+\phi_b^2)(M_x^2+M_y^2)+\no\\
&&g_1(\phi_a^2-\phi_b^2)(M_x^2-M_y^2)+2g_2\phi_a\phi_bM_xM_y\no\\ 
\eea
In Eqs.~(\ref{eq:fho1},\ref{eq:fho2}) we have $a(T)=a_0(T-T_h)$ and $a_0,b,b_{1,2} > 0$. Finally  the purely magnetic part of the free energy    given by 
\bea
F_m\{\vM\}&=&\frac{\vM^2}{2\chi_0}+\frac{1}{2}\lambda_1(M_x^4+M_y^4)+\lambda_2M_x^2M_y^2 -\vM\cdot\vH\no\\
\label{eq:fmu}
\eea
has to be added. Here $\vH =(H_x,H_y)$ and $\vM = (M_x,M_y)$ are external field and uniform magnetisation respectively lying in the tetragonal ab plane. The magnetization is determined by $\vM = \tensor{\chi}\vH$. In zero field the susceptibility tensor is diagonal with $\chi_{xx}=\chi_{yy}= \chi_0$ equal to the background susceptibility $\chi_0$ of \UR~ which is almost temperature independent for field in the plane \cite{Sugiyama99}. In an external field in general  $\Delta\chi =\chi_{xx}-\chi_{yy}$ and $\chi_{xy}, \chi_{yx}$ may become different from zero due the underlying hidden order.\\

At zero field minimization of the free energies for 1D representations leads to the order parameter ($T< T_h$)
\bea
\phi^2=\frac{a_0}{b}(T_h-T); \qquad F_0=-\frac{1}{2}\frac{a_0^2}{b}(T_h-T)^2
\label{eq:op1}
\eea
For degenerate E representation there are two possible solutions characterised by the vector  $\boldphi = (\phi_a,\phi_b)=\phi(\eta_a,\eta_b)$. Because $\boldphi$ is a commensurate ($\vQ=0$ or $(0,0,1)$) order parameter of multipolar density type the amplitudes $(\phi_a,\phi_b)$ may be chosen as real. The two possible phases $E(\eta_a,\eta_b)$ are then determined by the numbers $\eta_a=0,\pm 1$ and $\eta_b=0,\pm 1$ according to
\bea
&&E(0,1)\quad (b_1>b_2):\no\\ 
&&\phi^2=\frac{a_0}{b_1}(T_h-T); \; F_0=-\frac{1}{2}\frac{a_0^2}{b_1}(T_h-T)^2\\
&&E(1,1)\quad (b_1< b_2):\no\\ 
&&\phi^2=\frac{a_0}{b_1+b_2}(T_h-T); \;  F_0=-\frac{a_0^2}{b_1+b_2}(T_h-T)^2\no
\label{eq:op2}
\eea
All other combinations of $E(\eta_a,\eta_b)$ represent different domains of these phases. There are four domains with equal $F_0$ in each case:
$E(0,\pm 1)$, $E(\pm 1,0)$ for $b_1>b_2$ and $E(\pm 1,\pm 1)$ for   $b_1< b_2$.\\

For finite field the free energy functional has to be minimized simultaneously with respect to the HO parameters and the
induced uniform magnetization. After some lengthy but straightforward algebra the solution for all 1D and 2D representations may
be obtained. From this we calculate $\Delta\chi =\chi_{xx}-\chi_{yy}$ and $\chi_{xy}, \chi_{yx}$ as function of temperature and field for each symmetry which then determines the torque oscillations according to Eqs.~(\ref{eq:torque1},\ref{eq:torque2}). We give the set of results  for all HO symmetries of interest and then compare to experimental observations to draw conclusions on the prefered symmetry of HO.\\

The total torque in Eq.~(\ref{eq:torque1}) may always be split into contributions with twofold and fourfold oscillations
denoted by $\tau_\varphi^{(2)}$ and $\tau_\varphi^{(4)}$ respectively. The total $\tau(\varphi)$ may be written in reduced form as 
\begin{widetext}
\bea
\tau_\varphi^{(2)}+\tau_\varphi^{(4)}=
\tau(\varphi)&=&\tau_0(H)\Bigl[\frac{1}{2\chi_0}(\chi_{xy}-\chi_{yx})
+ \frac{\Delta\chi}{2\chi_0}\sin 2\varphi 
-\frac{1}{2\chi_0}(\chi_{xy}+\chi_{yx})\cos 2\varphi\Bigr]
\label{eq:torque2}
\eea
\end{widetext}
with the scale of the torque given by $\tau_0(H)=\mu_0VH^2\chi_0$. In this slightly more general form of Eq.~(\ref{eq:torque1}) it is taken into account that HO may possibly break reflection symmetry with respect to [110] diagonals in the plane, leading to $\chi_{xy}\neq\chi_{yx}$ for general field direction. Below we compile the results for the torque oscillations for the main candidates of HO symmetry. This is most conveniently done in a reference frame where the field direction refers to the crystal axis, i.e. ${\bf H}=H(\cos\varphi,\sin\varphi)$ with $\varphi$ counted from a-axis or [100] direction. Experimentally however the oscillations are preferably counted from the [110] direction i.e. with ${\bf H}=H(\cos(\hvarphi+\frac{\pi}{4}),\sin(\hvarphi+\frac{\pi}{4}))$ with $\varphi=\hvarphi+\frac{\pi}{4}$. For clarity we will give two- and fourfold oscillations for both conventions. They are simply related by using the identities $\sin 2\varphi=\cos 2\hvarphi,\; \cos 2\varphi = -\sin 2\hvarphi,\; \sin 4\varphi = -\sin 4\hvarphi,\; \cos 4\varphi = -\cos 4\hvarphi$. For  comparison with the experimentally observed oscillation pattern we will use the $\hvarphi$ convention .\\

First we discuss the results for the nondegenerate HO phases.
For convenience we introduce the constants $\delta_0=2\chi_0^2(\lambda_2-\lambda_1)$ and $\Delta_0=g\chi_0(a_0\tth/b)^\frac{1}{2}$. The fully symmetric terms in the Landau functional Eq.~(\ref{eq:fho1}) only lead to a shift of the HO transition temperature  in the field according to $\tilde{T}_h=T_h-\Delta T_h$ with  $\Delta T_h=(\tilde{g}\chi_0/a_0)(\chi_0H^2)$. Then we obtain for\\

$B_1 (x^2-y^2)$ quadrupolar HO:
\bea
\frac{\Delta\chi}{\chi_0}&=&-4\Delta_0\Bigl(1-\frac{T}{\tth}\Bigr)^\frac{1}{2}+\delta_0(\chi_0H^2)\cos 2\varphi\no\\
\frac{\chi_{xy}}{\chi_0}&=&0\no\\\\
\tau_\varphi^{(2)}&=&-2\tau_0\Delta_0(1-\frac{T}{\tth})^\frac{1}{2}\sin 2\varphi\no\\
\qquad \tau_\varphi^{(4)}&=&\frac{1}{4}\tau_0\delta_0(\chi_0H^2)\sin 4\varphi\no\\
\tau_{\hvarphi}^{(2)}&=&-2\tau_0\Delta_0(1-\frac{T}{\tth})^\frac{1}{2}\cos 2\hvarphi\no\\
\qquad \tau_{\hvarphi}^{(4)}&=&-\frac{1}{4}\tau_0\delta_0(\chi_0H^2)\sin 4\hvarphi\no
\label{eq:B1}
\eea

This model in principle leads to both twofold (period $\pi$) and fourfold (period $\pi/2$) torque oscillations with amplitudes $\tau_{\hvarphi}^{(2)}(H,T)$ and 
$\tau_{\hvarphi}^{(4)}(H,T)$ respectively. However for HO with $\vQ\neq 0$ as is presumably the case we have $g, \Delta_0 \equiv 0$ and therefore 
$\tau_{\hvarphi}^{(2)}(H,T)=0$ , i.e. no twofold oscillations appear.  Furthermore even for $\vQ =0$ when $\tau_{\hvarphi}^{(2)}(H,T)$ is  nonzero their angular dependence $\sim \cos 2\hvarphi $ has the wrong phase with maxima and minima interchanged compared to experimetal observation.
Therefore  $B_1 (x^2-y^2)$ type quadrupolar HO is not compatible with torque experiments.\\

$B_2 (xy)$ quadrupolar HO:
\bea
\frac{\Delta\chi}{\chi_0}&=&\delta_0(\chi_0H^2)\cos 2\varphi\no\\
\qquad \frac{\chi_{xy}}{\chi_0}&=&-\Delta_0\Bigl(1-\frac{T}{\tth}\Bigr)^\frac{1}{2}\no\\
\tau_\varphi^{(2)}&=&\tau_0\Delta_0(1-\frac{T}{\tth})^\frac{1}{2}\cos 2\varphi\no\\
\tau_\varphi^{(4)}&=&\frac{1}{4}\tau_0\delta_0(\chi_0H^2)\sin 4\varphi\\
\tau_{\hvarphi}^{(2)}&=&-\tau_0\Delta_0(1-\frac{T}{\tth})^\frac{1}{2}\sin 2\hvarphi\no\\
\tau_{\hvarphi}^{(4)}&=&-\frac{1}{4}\tau_0\delta_0(\chi_0H^2)\sin 4\hvarphi\no
\label{eq:B2}
\eea

This model also has both twofold and fourfold oscillations. Again for HO with $\vQ\neq 0$ the former vanish identically due to  $g, \Delta_0 \equiv 0$.
Furthermore for the case $\vQ =0$ where they are non-zero the amplitude increases  $\sim (T-\tth)^\frac{1}{2}$ as in the $B_1$ case. This is a much too strong increase  below $T_h$ in comparison with experiment that seem to follow a weaker dependence $\tau_{\hvarphi}^{(2)}(H,T)\sim c_1(T-\tth) + c_2(T-\tth)^2$. Therefore the $B_2(xy)$ HO model cannot explain the observed torque oscillations in \UR.\\

$A_2 (xy(x^2-y^2)$) hexadecapolar HO:
\bea
\frac{\Delta\chi}{\chi_0}&=&\delta_0(\chi_0H^2)\cos 2\varphi\no\\
\frac{1}{2\chi_0}(\chi_{xy}+\chi_{yx}))&=&-2(\chi_0\Delta_0)(\chi_0H^2)\Bigl(1-\frac{T}{\tth}\Bigr)^\frac{1}{2}\cos 2\varphi \no\\
\frac{1}{2\chi_0}(\chi_{xy}-\chi_{yx}))&=&-(\chi_0\Delta_0)(\chi_0H^2)\Bigl(1-\frac{T}{\tth}\Bigr)^\frac{1}{2}\no\\
\tau_\varphi^{(2)}&=&0\\
\tau_\varphi^{(4)}&=&\frac{1}{4}\tau_0(\chi_0H^2)\bigl[\delta_0\sin 4\varphi\no\\ 
&&+4(\chi_0\Delta_0)(1-\frac{T}{\tth}\Bigr)^\frac{1}{2}\cos 4\varphi]\no\\
\tau_{\hvarphi}^{(2)}&=&0\no\\
\tau_{\hvarphi}^{(4)}&=&-\frac{1}{4}\tau_0(\chi_0H^2)\bigl[\delta_0\sin 4\hvarphi\no\\ 
&&+4(\chi_0\Delta_0)(1-\frac{T}{\tth}\Bigr)^\frac{1}{2}\cos 4\hvarphi]\no
\label{eq:A2}
\eea

The $A_2$ symmetry is not realized for quadrupolar (rank 2) HO because it only breaks reflection symmetry
with respect to [110] type planes but not C$_4$ rotational symmetry. It appears first as rank 4 (hexadecapolar) HO parameter.
However in this case $\tau_{\hvarphi}^{(2)}\equiv 0$, independent of \vQ~,  in conflict with experiment. Furthermore for $\vQ = 0$ $(\Delta_0\neq 0)$
the fourfold oscillations below $T_h$ would be skewed with respect to the [110] diagonals due to the abovementioned lack of reflection symmetry
of the hexadecapolar HO. This feature is also incompatible with torque experiments. We conclude that the hexadecapolar order parameter 
cannot explain the observed twofold torque oscillations.\\

In summary if we restrict to considering only $\vQ\neq 0$ HO then none of the 1D representations can produce the twofold torque or susceptibility oscillations that have been observed by Okazaki et al \cite{Okazaki11} below $T_h$. For $\vQ = 0$ $B_1$ and $B_2$ quadrupolar HO may lead to twofold oscillations but they have the wrong temperature behaviour or angular dependence and therefore are not compatible with experiments. Furthermore the $A_2$ hexadecapolar OP will never lead to the observed twofold torque oscillations.
The failure of all 1D HO parameter models cannot be remedied by including higher order terms (in $\phi$, \vM) in the Landau functionals. They would only renormalize the amplitudes but would not change their principal characteristics.\\

Therefore we are lead to consider the twofold degenerate E representation with its two different phases as only remaining candidates. In this case the fully symmetric terms
in Eq.~(\ref{eq:fho2}) lead to a shifted transition temperature $\tilde{T}_h=T_h-\Delta T_h$ with  $\Delta T_h=(\tilde{g_0}\chi_0/a_0)(\chi_0H^2)$. Then we obtain for\\

$E(0,1) (0,zx) $ quadrupolar HO:\\
\bea
\label{eq:E10}
\frac{\Delta\chi}{\chi_0}&=&4\Delta_0(1-\frac{T}{\tth})+ \tilde{\delta}_0(\chi_0H^2)\cos 2\varphi\no\\
\frac{\chi_{xy}}{\chi_0}&=&0\no\\
\Delta_0&=&(g_1\chi_0)(a_0\tth/b_1); \;
\tilde{\delta}_0=2\chi_0^2\Bigl[(\lambda_2-\lambda_1)+\frac{2g_0^2}{b_1}\Bigr]\no \\
\tau_\varphi^{(2)}&=&2\tau_0\Delta_0(1-\frac{T}{\tth})\sin 2\varphi\\
\tau_\varphi^{(4)}&=&\frac{1}{4}\tau_0\tilde{\delta}_0(\chi_0H^2)\sin 4\varphi\no\\
\tau_{\hvarphi}^{(2)}&=&2\tau_0\Delta_0(1-\frac{T}{\tth})\cos 2\hvarphi\no\\
\tau_{\hvarphi}^{(4)}&=&-\frac{1}{4}\tau_0\tilde{\delta}_0(\chi_0H^2)\sin 4\hvarphi\no
\eea

The expressions for the $E(1,0)$ domain may be obtained by substituting 
$\Delta_0\rightarrow -\Delta_0$ in the above expressions. It means that twofold amplitude changes sign
$\tau_{\hvarphi}^{(2)}\rightarrow -\tau_{\hvarphi}^{(2)}$ when we change from E(0,1) to E(1,0) domain.\\

This is the first model which leads to twofold oscillations $\tau_{\hvarphi}^{(2)}\neq 0$ even in the case $\vQ\neq 0$. This is because $g_1$ may be finite also for finite \vQ~ since the quadratic term $\sim (\phi_a^2-\phi_b^2)$ in Eq.~(\ref{eq:fho2}) is translationally invariant. However  $\tau_{\hvarphi}^{(2)}\sim\cos 2\hvarphi$ and therefore its maxima/minima are shifted by $\pi/4$ with respect to the correct experimental positions. Due to these wrong angular characteristics the E(0,1) or E(1,0) phase can also be excluded as the HO candidate. Finally we discuss\\

$E(1,1) (yz,zx) $ quadrupolar HO:\\
\bea
\label{eq:E11}
\frac{\Delta\chi}{\chi_0}&=&\tilde{\delta}_0(\chi_0H^2)\cos 2\varphi\no\\
\frac{\chi_{xy}}{\chi_0}&=&-2\Delta_0\Bigl(1-\frac{T}{\tth}\Bigr)\no\\
\Delta_0&=&g_2\chi_0(a_0\tth/(b_1+b_2))\no\\
\qquad \tilde{\delta}_0&=&2\chi_0^2\Bigl[(\lambda_2-\lambda_1)+\frac{4g_1^2}{b_1-b_2}\Bigr] \no\\
\tau_\varphi^{(2)}&=&2\tau_0\Delta_0(1-\frac{T}{\tth})\cos 2\varphi \\
\qquad \tau_\varphi^{(4)}&=&\frac{1}{4}\tau_0\tilde{\delta}_0(\chi_0H^2)\sin 4\varphi\no\\
\tau_{\hvarphi}^{(2)}&=&-2\tau_0\Delta_0(1-\frac{T}{\tth})\sin 2\hvarphi\no\\
\tau_{\hvarphi}^{(4)}&=&-\frac{1}{4}\tau_0\tilde{\delta}_0(\chi_0H^2)\sin 4\hvarphi\no
\eea
%

%
\begin{figure*}
\vspace{0.2cm}

\includegraphics[width=120mm]{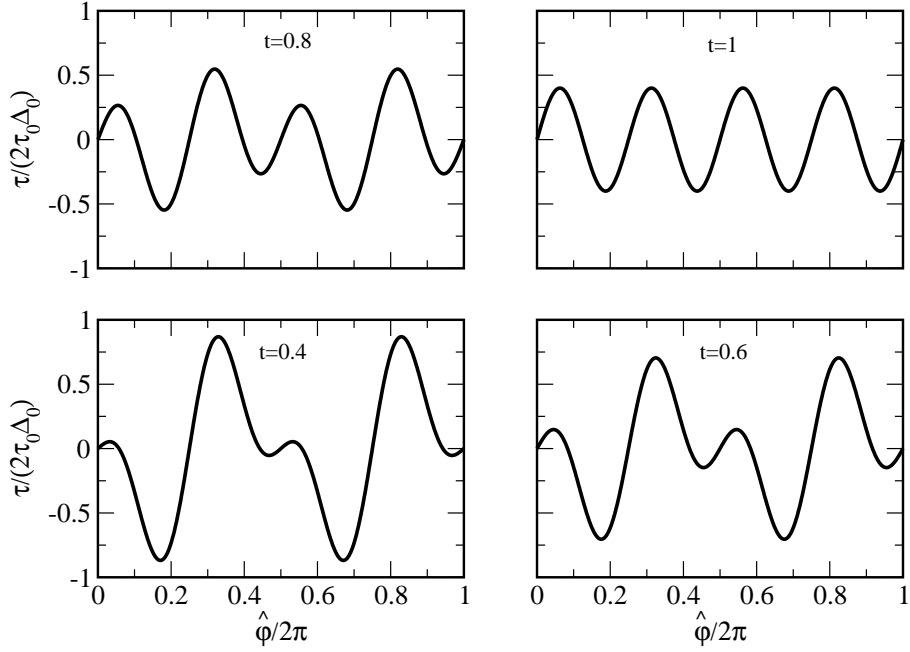}
\caption{Torque oscillations in the E(1,1) phase as function of in-plane field angle $\hvarphi=\varphi-\frac{\pi}{4}$ (with respect to [110] direction) for several
temperatures $t=T/\tth$. Above and at $\tth$ only the fourfold oscillations due to magnetic anisotropy are present. Below $\tth$ the twofold oscillations
resulting from the E(1,1)- type hidden order appear and increase with lowering the temperature. Plots have been made for an amplitude ratio $\tilde{\delta}_0(\chi_0H^2)/(8\Delta_0)  = -0.4$.}
\label{fig:Fig1}
\end{figure*}
%

This model has the proper behaviour of torque angular dependence $\tau_{\hvarphi}^{(2)}, \tau_{\hvarphi}^{(4)}$  in comparison to experiment as shown in Fig.~\ref{fig:Fig1}. The fourfold oscillation is present already above $\tth$ and temperature independent in the leading order. On the other hand the twofold oscillation sets in below $\tth$ and grows linearly with $(1-T/\tth)$ in leading order. If we also include terms $\sim \tilde{g}_1(\phi_a^4-\phi_b^4)(M_x^2-M_y^2)$ the temperature dependence will be given by  $\tau_{\hvarphi}^{(2)}\sim\Delta_0(1-\frac{T}{\tth})+\tilde{\Delta}_0(1-\frac{T}{\tth})^2$. It appears that in experiment both terms are present, at least in the samples with small size. Experimentally it was also found that $\tau_{\hvarphi}^{(2)},
\tau_{\hvarphi}^{(4)}$ have opposite signs which implies $\lambda_2 <\lambda_1$ and $\tilde{\delta}_0<0$.
We conclude that the basic features of the torque experiment in Ref.~  \cite{Okazaki11} are reproduced by the E(1,1) quadrupolar HO model. This is in agreement with the order parameter proposed from analysis of resistivity anisotropy in Ref.~\onlinecite{Miyake10}.
As explained above there are four domains with $(\eta_a,\eta_b) =(\pm 1,\pm 1)$ in this phase. The domain type matters only in the last term of Eq.~(\ref{eq:fho2}) through the sign of $\eta_a\eta_b =\pm 1$. For domains with  $\eta_a\eta_b = -1$ one has to replace $\Delta_0\rightarrow -\Delta_0$ in Eq.~(\ref{eq:E11}) leading to opposite sign for the twofold oscillations. If domains with $\eta_a\eta_b=\pm 1$ are equally distributed their twofold amplitudes will cancel. Indeed the latter are most prominent in the smallest samples where one would expect a single domain phase like $E(1,1)$ to be realised. Finally we may rewrite the twofold oscillation in Eq.~(\ref{eq:E11}) in the form 
\bea
\tau_{\hvarphi}^{(2)}&=&A_{\hvarphi}^{(2)}\sin 2\hvarphi\no\\
\frac{A_{\hvarphi}^{(2)}}{(\mu_0H)}&=&V\bigl(\frac{\chi_{xy}}{\chi_0}\bigr)M_0
= \frac{V}{2}[M_{[110]}-M_{[\bar{1}10]}]
\label{eq:E11v2}
\eea
where $M_0=\chi_0H$. This shows explicitly that the twofold amplitude in the E(1,1) phase is due to the  off-diagonal susceptibility induced by the background HO. The last identity may be obtained from $M(\varphi)\simeq (\chi_0 H)[1+(\chi_{xy}/\chi_0)\sin 2\varphi]$. It signifies directly the breaking of fourfold symmetry in the HO phase with respect to the diagonal $[100]$ and $[\bar{1}10]$ axes.
The linear relation in Eq.~(\ref{eq:E11v2}) is found experimentally for low fields. 
For larger fields there is however a sign reversal. Since the latter depends on the type of domain we speculate that this may be due to change of domain preference at higher fields.

\section{Classification of possible field induced dipole moments}
\label{sect:dipole}

\begin{table}[t]
\begin{center}
\begin{tabular}{ccccc}
\hline
$\Gamma$(C$_{\rm 4v}$) & $O^{\Gamma^+}$ & $\phi^{\Gamma^+}_{n}$(Even) 
 & $O^{\Gamma^-}$ & $\phi^{\Gamma^-}_{n}$(Odd)\\
\hline
$\Gamma_1$ & $N_f$ & $\phi^{A^{+}_{1}}_{1}=\rho_{11}$ 
   & $J^z$ & $\phi^{A^{-}_{2}}_{1}=S^{z}_{11}$\\
\hline
$\Gamma_3$ & $O_2^2$ 
  & $\phi^{B^{+}_{1}}_{1}=\frac{1}{\sqrt{2}}(\rho_{23}+\rho_{32})$ 
   & $T^{\beta}_z$ 
    & $\phi^{B^{-}_{2}}_{1}=\frac{1}{\sqrt{2}}(S^{z}_{23}+S^{z}_{32})$\\
\hline
$\Gamma_4$ & $O_{xy}$ 
  & $\phi^{B^{+}_{2}}_{1}=\frac{i}{\sqrt{2}}(S^{z}_{23}-S^{z}_{32})$ 
   & $T_{xyz}$ 
    & $\phi^{B^{-}_{1}}_{1}=\frac{i}{\sqrt{2}}(\rho_{23}-\rho_{32})$\\
\hline
$\Gamma_5$ & $O_{yz}$ 
  & $\phi^{E^{+}}_{1x}=\frac{i}{\sqrt{2}}(S^{x}_{12}-S^{x}_{21})$ 
   & $J_x$ 
    & $\phi^{E^{-}}_{1x}=S^x_{11}$\\
 & $H^{\alpha}_x$ 
  & $\phi^{E^{+}}_{2x}=\frac{i}{\sqrt{2}}(S^{x}_{23}-S^{x}_{32})$ 
   & $T^{\alpha}_x$ 
    & $\phi^{E^{-}}_{2x}=S^x_{22}$\\
 & $H^{\beta}_x$ 
  & $\phi^{E^{+}}_{3x}=\frac{i}{\sqrt{2}}(S^{x}_{31}-S^{x}_{13})$ 
   & $T^{\beta}_x$ 
    & $\phi^{E^{-}}_{3x}=S^x_{33}$\\
\cline{2-5}
 & $O_{zx}$ 
  & $\phi^{E^{+}}_{1y}=\frac{i}{\sqrt{2}}(S^{y}_{12}-S^{y}_{21})$ 
   & $J_y$ 
    & $\phi^{E^{-}}_{1y}=S^y_{11}$\\
 & $H^{\alpha}_y$ 
  & $\phi^{E^{+}}_{2y}=\frac{i}{\sqrt{2}}(S^{y}_{23}-S^{y}_{32})$ 
   & $T^{\alpha}_y$ 
    & $\phi^{E^{-}}_{2y}=S^y_{22}$\\
 & $H^{\beta}_y$ 
  & $\phi^{E^{+}}_{3y}=\frac{i}{\sqrt{2}}(S^{y}_{31}-S^{y}_{13})$ 
   & $T^{\beta}_y$ 
    & $\phi^{E^{-}}_{3y}=S^y_{33}$\\
\hline
\end{tabular}
\end{center}
\caption{The first column shows 
irreducible representations of C$_{4v}$ for
magnetic field along [001] direction. 
The second and third ones describe corresponding bases 
consisting of multipole moments and 
one-particle operators with even parity, 
respectively.  Those with odd parity are given in
the fourth and fifth columns.}
\label{table:table3}
\end{table}

We have shown that the  results of torque experiments  and their Landau theory analysis strongly favor the antiferroquadrupolar HO with E(xz,yz) symmetry of the (1,1) type and wave vector $\vQ=(0,0,1)$. One of the most direct ways to look for hidden order is to identify the direction of induced staggered dipolar moments in an external field. For field along [001] or [100] directions the symmetry is reduced from D$_{4h}$ to C$_{4v}$ or C$_{2v}$ respectively. The  proposed candidates for primary underlying hidden order (quadrupolar or hexadecapolar)  have all even time reversal parity. Then a magnetic field may induce dipolar or octupolar moments which have odd time reversal parity. The full compatibility tables for the induced moments
and field along the two symmetry directions are given in Tables~\ref{table:table3} and \ref{table:table4}. Note that more cases of HO symmetry have been included in the tables as compared to the previous Landau analysis. Table ~\ref{table:table3}  shows that for field along [001] the E(1,1) type quadrupolar HO may have induced moments along the [110] diagonal directions. It is also obvious from Table ~\ref{table:table4} that for quadrupolar E (0,1) ($O_{zx}$) type HO  and field along [100] the induced staggered moments $\vm_\vQ$ should be oriented along [001] direction. The same holds true for E(1,0) ($O_{yz}$) and field along [010]. If the field is perpendicular to the domain orientation of HO no dipolar moment but only $T_{xyz}$ octupolar moments may be induced. For the E(1,1) phase both quadrupolar components lead to an induced moment along [001]. This is indeed the same direction  as observed in the genuine AF phase of \UR~ above the critical pressure of 0.7 GPa \cite{Amitsuka07}. The moment dependence for general field direction in the ab-plane may be calculated from an extended Landau functional including the staggered magnetisation ($m_{z\vQ}$) terms. It is given by
\bea
F&=&F_{HO}(\phi_a,\phi_b)+F_m(\vM)
+\frac{1}{2\chi_\vQ}\vm_\vQ^2 \no\\
&&+\lambda_\vQ(H_xm_{z\vQ}\phi_a+H_ym_{z\vQ}\phi_b)
\label{eq:fmst}
\eea
Here $\chi_\vQ$ denotes the staggered susceptibility and $\lambda_\vQ$ a dimensionless coupling constant.
Mininimization of F leads to the expression for $m_{z\vQ}$. The physical important quantity is the intensity of field induced Bragg peaks which is $\sim |m_{z\vQ}|^2$. It is given by the following domain dependent expressions
\bea
E(1,\pm 1):\quad |m_{z\vQ}|^2 &=&(\lambda_\vQ\chi_\vQ)^2H^2(1\pm\cos 2\hvarphi)\phi^2\no\\
E(1,0):\quad |m_{z\vQ}|^2 &=&(\lambda_\vQ\chi_\vQ)^2H^2\frac{1}{2}(1-\sin 2\hvarphi)\phi_a^2\no\\
E(0,1):\quad |m_{z\vQ}|^2 &=&(\lambda_\vQ\chi_\vQ)^2H^2\frac{1}{2}(1+\sin 2\hvarphi)\phi_b^2\no\\
\label{eq:Est}
\eea
where  $\phi^2=(a_0/(b_1+b_2))(\tth-T)$ and   $\phi_{a,b}^2=(a_0/b_1)(\tth-T)$ .
As most promising candidate for HO the E(1,1) phase has been identified.
In a diffraction experiment the intensities at \vQ~ and equivalent points should vary according to the first of the above expressions. As for the torque, twofold oscillations as function of in-plane field angle in the intensity are predicted for a single domain. If domains with $\eta_a\eta_b=\pm 1$ are equally populated
only the average intensity proportional to  $ |m_{z\vQ}|^2 = (\lambda_\vQ\chi_\vQ)^2H^2\phi^2$  should be observed which does not depend on the field angle.
As in the case of torque experiments, twofold angular variation of Bragg intensities should therefore only be observed in small single domain samples.  Here $\chi_\vQ$ can be expected to be of order $1/T_h$. Then $(H/T_h)^2$ sets the scale for the field induced moment. Since $T_h =18 $ K is quite large one may conjecture that magnetic moments induced out of the HO phase appear only at very large fields. This means the intensity of magnetic  Bragg peaks at accessible fields is likely to be dominated by those of the tiny volume fraction ($\sim 1\%$ according to Ref. \onlinecite{Yokoyama05}) of the parasitic AF phase which are already present at zero field \cite{Mason95}.

\begin{table}[t]
\begin{center}
\begin{tabular}{ccccc}
\hline
$\Gamma$(C$_{\rm 2v}$) & $O^{\Gamma^+}$ & $\phi^{\Gamma^+}_{n}$(Even) 
 & $O^{\Gamma^-}$ & $\phi^{\Gamma^-}_{n}$(Odd)\\
\hline
$\Gamma_1$ & $N_f$ & $\phi^{A^{+}_{1}}_{1}=\rho_{11}$ 
   & $J_x$ 
    & $\phi^{E^{-}}_{1x}=S^x_{11}$\\
 & $O_2^0$ & $\phi^{A^{+}_{1}}_{2}=\rho_{22}$ 
   & $T^{\alpha}_x$ 
    & $\phi^{E^{-}}_{2x}=S^x_{22}$\\
 & $O_2^2$ & $\phi^{A^{+}_{1}}_{3}=\rho_{33}$ 
   & $T^{\beta}_x$ 
    & $\phi^{E^{-}}_{3x}=S^x_{33}$\\
\hline
$\Gamma_2$ & $O_{yz}$ 
  & $\phi^{E^{+}}_{1x}=\frac{i}{\sqrt{2}}(S^{x}_{12}-S^{x}_{21})$ 
   & $T_{xyz}$ 
    & $\phi^{A^{-}_{1}}_{1}=\frac{i}{\sqrt{2}}(\rho_{12}-\rho_{21})$\\
\hline
$\Gamma_3$ & $O_{xy}$ 
  & $\phi^{A^{+}_{2}}_{1}=\frac{i}{\sqrt{2}}(S^{z}_{12}-S^{z}_{21})$
   & $J_y$ 
    & $\phi^{E^{-}}_{1y}=S^y_{11}$\\
 & $H^{\alpha}_z$ 
  & $\phi^{B^{+}_{2}}_{1}=\frac{i}{\sqrt{2}}(S^{z}_{23}-S^{z}_{32})$ 
   & $T^{\alpha}_y$ 
    & $\phi^{E^{-}}_{2y}=S^y_{22}$\\
 & $H^{\beta}_z$ 
  & $\phi^{B^{+}_{2}}_{2}=\frac{i}{\sqrt{2}}(S^{z}_{31}-S^{z}_{13})$ 
   & $T^{\beta}_y$ 
    & $\phi^{E^{-}}_{3y}=S^y_{33}$\\
\hline
$\Gamma_4$ & $O_{zx}$ 
  & $\phi^{E^{+}}_{1y}=\frac{i}{\sqrt{2}}(S^{y}_{12}-S^{y}_{21})$ 
   & $J^z$ & $\phi^{A^{-}_{2}}_{1}=S^{z}_{11}$\\
 & $H^{\alpha}_y$ 
  & $\phi^{E^{+}}_{2y}=\frac{i}{\sqrt{2}}(S^{y}_{23}-S^{y}_{32})$ 
   & $T^{\alpha}_z$ & $\phi^{A^{-}_{2}}_{2}=S^{z}_{22}$\\
 & $H^{\beta}_y$ 
  & $\phi^{E^{+}}_{3y}=\frac{i}{\sqrt{2}}(S^{y}_{31}-S^{y}_{13})$ 
   & $T^{\beta}_z$ & $\phi^{A^{-}_{2}}_{3}=S^{z}_{33}$\\
\hline
\end{tabular}
\end{center}
\caption{The first column shows 
irreducible representations of C$_{2v}$ for
magnetic field along [100] direction. 
The second and third ones describe corresponding bases 
consisting of multipole moments and 
one-particle operators with even parity, 
respectively. Those with odd parity are given
in the fourth and fifth columns.}
\label{table:table4}
\end{table}

\section{Elastic constant anomalies below the HO transiton}
\label{sect:elastic}

As mentioned in the introduction the absence of any homogeneous lattice distortions below $T_h$  \cite{Jeffries10} suggest HO at nonzero wave vector. On the other hand small but distinct anomalies in the elastic constants below the HO transition of \UR~ have been found \cite{Kuwahara97}. 
In the same manner as torque experiments probe the magnetic (dipolar) susceptibilities the elastic constant anomalies probe the quadrupolar susceptibilities associated with background HO  below $T_h$. As in the torque experiments for $\tensor{\chi}$ in Ref.~\onlinecite{Okazaki11} the ultrasonic experiments for the symmetry elastic constants $c_\Gamma$  \cite{Kuwahara97} are sensitive for the changes caused by the onset of hidden order. The results obtained in Ref.~\cite{Kuwahara97} have, however, sofar not been analyzed in their relevance for the HO symmetry questions, only the much larger anomalies due to local single ion quadrupolar CEF excitations have been disussed.
Generally it was found that elastic constant anomalies $\Delta c_\Gamma$ below $T_h$ are quite small of the order of $\Delta c_\Gamma/c_\Gamma\simeq 10^{-5}$ due to a slight change of slope in $c_\Gamma(T)$. It is not clear at present whether this change is of significance and contains information on the symmetry of HO. Therefore we investigate this problem in the present phenomenological context.\\

\begin{table*}[tb]
\caption{Step-like elastic constant anomalies due to linear strain coupling $F^{(1)}_{st-HO}$  in the E type HO phase for transverse $c_{11}-c_{12}$ and $2c_{66}$ symmetry. Note that the anomaly is absent for the prefered HO symmetry E(1,1)  which is well compatible with experiments \cite{Kuwahara97}.
Here $\Delta c_\gamma$ and $\Delta c_\delta$ are independent of domain type in both phases.}
\begin{center}
\begin{tabular}{cccc}
\hline
\\[0.05cm]
phase \qquad& $ \Delta c_\gamma (c_{11}-c_{12})$\qquad & $\Delta c_\delta (2c_{66}) $  \qquad& OP \qquad \\
\\[0.05cm]
\hline
\hline
\\[0.1cm]
$E(0,1), E(1,0)$& $-\frac{g_\gamma^2}{b_1}(1+2\frac{\tilde{g}_\gamma}{g_\gamma}\phi^2)^2 < 0$ 
   &$ \frac{g_\delta^2}{2(b_1-b_2)} > 0$
    & $\phi^2=\frac{a}{b_1}; (b_1>b_2)$\\
\\[0.1cm]    
\hline
\\[0.1cm]
$E(1,1)$\;\;\;\;& $\frac{2g_\gamma^2}{(b_2-b_2)}(1+2\frac{\tilde{g}_\gamma}{g_\gamma}\phi^2)^2 > 0$\;\;\;\;
   & 0 \;\;\;\;   
   &  $\phi^2=\frac{a}{b_1+b_2}; (b_1< b_2)$\;\;\;\;  \\ 
\\[0.1cm]
\hline
\end{tabular}
\end{center}
\label{table:table2}
\end{table*}
%

To perform such analysis we extend the previous Landau treatment  for zero field and include the coupling terms to the homogeneous elastic strains. They can be classified  (e.g.,  in Ref.\onlinecite{Kuwahara97}) into the $D_{4h}$ symmetry strains $\epsilon_\Gamma$ associated with elastic constants $c_\Gamma$ leading to the elastic energy
\bea
F_{el}= \frac{1}{2} \sum_\Gamma c_\Gamma^0\epsilon_\Gamma^2
\label{eq:fel}
\eea
Here $c_\Gamma^0(T)$ is the elastic background constant in quasiharmonic approximation (containing the anharmonic contribution and the effect of localised CEF excitations). The total elastic constant $c_\Gamma(T)=c_\Gamma^0(T)+\Delta c_\Gamma(T)$ contains the small effect of the coupling to HO below T$_h$ and this part may in principle give additional information on the  symmetry of the latter.\\

For $D_{4h}$ symmetry there are two one dimensional $A_1$ type fully symmetric volume strains and in-plane tetragonal strains of $B_1$- type $\epsilon_\gamma=(1/\sqrt{2})(\epsilon_{xx}-\epsilon_{yy})$ with $c_\gamma=c_{11}-c_{12}$ and $B_2$- type  $\epsilon_\delta=\sqrt{2}\epsilon_{xy}$ with $c_\delta=2c_{66}$. In addition the two dimensional out-of plane strain of E-type is given by $\epsilon_\epsilon=\sqrt{2}(\epsilon_{zx},\epsilon_{yz})$ with $c_\epsilon$=2c$_{44}$. The symmetry strains couple to the HO parameter via i) first order (linear in $\epsilon_\Gamma$) strain coupling leading to corrections $\Delta c_\Gamma$ in second order of the coupling constant and ii) second order  (quadratic in $\epsilon_\Gamma$) leading to corrections $\Delta c'_\Gamma$ in first order of the coupling constants.\\

We first focus on the linear strain coupling.
For HO with $\vQ\neq 0$ translational invariance is broken and only the square (or generally even powers) of the nontrivial 1D order parameter can couple linearly to the 1D homogeneous strains. But for any 1D representation $\Gamma$ we have $\Gamma\times\Gamma=A_1$. Therefore the linear strain coupling of the order parameter can exist only for the volume strain . This means that the nontrivial elastic constants (aside from the bulk modulus) should not exhibit an anomaly at $T_h$ for $B_1$, $B_2$ and $A_2$ type HO resulting from linear strain coupling. Then the only case that remains to be discussed are the two possible E phases. The linear strain order-parameter free energy for this case is
\bea
F^{(1)}_{st-HO}&=& g_\gamma\epsilon_\gamma(\phi_a^2-\phi_b^2)+\tilde{g}_\gamma\epsilon_\gamma(\phi_a^4-\phi_b^4)
+g_\delta\epsilon_\delta\phi_a\phi_b \no\\
&&+\tilde{g}_\delta\epsilon_\delta(\phi_a^2+\phi_b^2)\phi_a\phi_b
\label{eq:stho}
\eea
Here we did not include the out-of plane E-type $c_{44}$ strain $\epsilon_\epsilon$ because it has no linear coupling when ${\bf Q}\neq 0$. The anomaly of the elastic constants below $T_h$ may now be obtained from
\bea
\Delta c_\Gamma(T)=
\sum_\alpha\phi'_{\Gamma\alpha}\Bigl(\frac{\partial^2F}{\partial\phi_\alpha\partial\epsilon_\Gamma}\Bigr);
\qquad \phi'_{\Gamma\alpha}=\frac{\partial\phi_\alpha}{\partial\epsilon_\Gamma}
\label{eq:cel}
\eea
where $F=F_0+F_{el}+F^{(1)}_{st-HO}$ and $\Gamma=\gamma,\delta$ denote $c_{11}-c_{12}$ and $c_{66}$ elastic constants, respectively. After determining $\phi'_{\Gamma,\alpha}$ from the equilibrium equations the calculation of $\Delta c_\Gamma$ is straightforward. The final results are shown in Table~\ref{table:table2} for the two possible E phases. Generally step-like $\Delta c_\Gamma(T)$$\sim g_\Gamma^2$ anomalies at T$_h$ are predicted. However as mentioned before experimentally rather slope changes in $c_\Gamma(T)$ are observed. This may be partly due to broadening of the transition as in the case of the specific heat anomaly at T$_h$. On the other hand slope changes are naturally obtained as a result of the quadratic strain coupling which we will discuss now. For the E-type HO this mechanism is described by 
\bea
F^{(2)}_{st-HO}=g'_\Gamma\epsilon_\gamma^2(\phi_a^2+\phi_b^2)+g'_\delta(\phi_a^2+\phi_b^2)
\label{eq:stho2}
\eea
where $g'_\Gamma$ are the second order strain coupling constants. Contrary to first order there is also a possible second order coupling of $\epsilon_\epsilon$ strains ($c_{44}$ elastic constant) but it will not be considered here. Because $\epsilon_\Gamma^2$ ($\Gamma=\gamma,\delta$) transforms like $A_1$ only coupling to the modulus of the order parameter is possible. Considering the form of $F_{el}$ the elastic constant corrections due to $F^{(2)}_{st-HO}$
can be read off immediately as
\bea
\Delta c'_\Gamma=\Delta c^{0'}_\Gamma(1-\frac{T}{T_h})\no\\
\label{eq:cel2}
\eea
therefore the anomalies due to quadratic strain coupling are continuous consisting in a slope change  below $T_h$ in contrast to the discontinuous jumps caused by linear strain coupling. The amplitudes $\Delta^{0'}_\Gamma$ are given by  domain independent expressions for the two E phases which are now linear in the coupling constants $g'_\Gamma$:
\bea
E(1,0), E(0,1):\quad \Delta c^{0'}_\Gamma&=&2g'_\Gamma\Bigl(\frac{a_0T_h}{b_1}\Bigr)\no\\
E(1,1):\quad \Delta c^{0'}_\Gamma&=&4g'_\Gamma\Bigl(\frac{a_0T_h}{b_1+b_2}\Bigr)
\label{eq:camp2}
\eea
Since only the (tiny) slope changes are observed in $c_\Gamma(T)$ in Ref.~\onlinecite{Kuwahara97} this may well be the dominant mechanism. We note however that a similar coupling as in Eq.~(\ref{eq:stho2}) exists for the nondegenerate $A_2, B_1, B_2$ HO phases. Simply $(\phi_a^2+\phi_b^2)$ has to be replaced by $\phi^2$ and the slope change anomalies are the same as in Eqs.~(\ref{eq:cel2},\ref{eq:camp2}) except for an overall factor of two. Therefore we conclude that the elastic constant anomalies of the type observed in  Ref.~\onlinecite{Kuwahara97} may not be very useful for discrimination between the possible symmetries of HO. On the other hand these measurements were still done in samples with parasitic small moment AF phase which may obscure the results. Therefore they should be repeated using the pure HO samples.

\section{conclusion}
\label{sect:conclusion}

In this work we have reconsidered the question of proper symmetry of the hidden order in \UR.
A phenomenological Landau analysis based on  $D_{4h}$ irreducible HO parameters has been applied to recent oscillatory torque experiments in rotating magnetic field. We have concluded that the experimental observations of twofold torque oscillations are only compatible with theoretical predictions from the two component (antiferro-) quadrupolar E(1,1) ($O_{yz}$,$O_{zx}$) phase . The second single component phase  E(1,0)  can be ruled out because of a shifted angular and wrong temperature dependence of the twofold amplitude. For single domain samples of the  E(1,1) phase exhibits twofold oscillations with the proper field angle and temperature dependence.

All one dimensional representations, namely quadrupolar $B_1(O_{x^2-y^2})$ and $B_2(O_{xy})$ as well as hexadecapolar $A_2(H_{xy(x^2-y^2)})$ are incompatible with torque experiments. We conclude that the two component E-type quadrupole is the HO symmetry in \UR. This order parameter is also compatible with observations in uniaxial stress experiments \cite{Yokoyama05} and resistivity anisotropy \cite{Miyake10}. The elastic constant anomalies observed sofar are possible in the E type hidden order but may presently not be used for discrimination of HO symmetry.

It remains to be seen whether the quadrupolar E(1,1) type HO can be identified directly. In this respect resonant x-ray scattering is a powerful method. Results obtained before \cite{Amitsuka10} exclude the $B_1(O_{x^2-y^2})$ and $B_2(O_{xy})$ type quadrupoles in agreement with present torque results. However possibility of quadrupolar E(1,1) type HO has sofar not been investigated by this method. One should be aware that due to the domain problem of the degenerate quadrupole it could only be used for small single domain samples.

\section*{Acknowledgements}
The authors would like to thank Takasada Shibauchi and Yuji Matsuda for communicating their experimental results before publication and for helpful discussions. The authors would also like to acknowledge hospitality of Ting-kuo Lee at the Academia Sinica in Taipei where part of this work has been performed.

\end{document}